\documentclass[twocolumn]{pasj00}
\draft
\usepackage[dvips]{color}
\usepackage{multicol}

\begin{document}
\SetRunningHead{N. Narita et al.}{The Rossiter-McLaughlin Effect of HD~17156}
\Received{2009/04/27}
\Accepted{2009/05/28}

\title{Improved Measurement of the
Rossiter-McLaughlin Effect in the Exoplanetary System HD~17156$^*$}


\author{
Norio \textsc{Narita},\altaffilmark{1}
Teruyuki \textsc{Hirano},\altaffilmark{2}
Bun'ei \textsc{Sato},\altaffilmark{3}
Joshua N.\ \textsc{Winn},\altaffilmark{4}
Yasushi \textsc{Suto},\altaffilmark{2}\\
Edwin L.\ \textsc{Turner},\altaffilmark{5,6}
Wako \textsc{Aoki},\altaffilmark{1}
Motohide \textsc{Tamura},\altaffilmark{1}
and Toru \textsc{Yamada}\altaffilmark{7}
}

\altaffiltext{1}{
National Astronomical Observatory of Japan, 2-21-1 Osawa,
Mitaka, Tokyo, 181-8588, Japan
}
\email{norio.narita@nao.ac.jp}

\altaffiltext{2}{
Department of Physics, The University of Tokyo, Tokyo, 113--0033, Japan
}

\altaffiltext{3}{
Global Edge Institute, Tokyo Institute of Technology,
2-12-1 Ookayama, Meguro, Tokyo, 152-8550, Japan
}

\altaffiltext{4}{
Department of Physics, and Kavli Institute for Astrophysics
and Space Research,\\
Massachusetts Institute of Technology, Cambridge, MA 02139, USA
}

\altaffiltext{5}{
Princeton University Observatory, Peyton Hall,
Princeton, NJ 08544, USA
}

\altaffiltext{6}{
Institute for the Physics and Mathematics of the Universe,
The University of Tokyo, Kashiwa, 277-8568, Japan
}

\altaffiltext{7}{
Astronomical Institute, Tohoku University, Aramaki, Aoba, Sendai,
980-8578, Japan
}


\maketitle

\begin{abstract}
  We present an improved measurement of the Rossiter-McLaughlin effect
  for the exoplanetary system HD~17156, based on radial-velocity data
  gathered with the Subaru 8.2m telescope throughout the planetary
  transit of UT~2008~November~7.  The data allow for a precise and
  independent determination of the projected spin-orbit angle of this
  system: $\lambda = 10.0^{\circ} \pm 5.1^{\circ}$.  This result
  supersedes the previous claim of $\lambda = 62^{\circ} \pm 25^{\circ}$
  by Narita et al., which was based on lower-precision data with poor
  statistics.  Thus the stellar spin and planetary orbital axes of the
  HD~17156 system are likely to be well-aligned, despite the planet's
  large orbital eccentricity suggesting a history of strong dynamical
  interactions.
\end{abstract}
\footnotetext[*]{Based on data collected at Subaru Telescope,
which is operated by the National Astronomical Observatory of Japan.}

\section{Introduction}

The discoveries of more than 340 extrasolar planets have revealed the
remarkable diversity of planetary systems. For example, many Jovian
planets lie far inside the snow line, and many of their orbits are more
eccentric than the orbits in the Solar system.  This diversity of
planetary systems has prompted theorists to consider new modes of planet
formation and migration, including Type II migration models (e.g.,
\cite{1985prpl.conf..981L, 1996Natur.380..606L, 2004ApJ...616..567I}),
planet-planet scattering models (e.g., \cite{1996Sci...274..954R,
2002Icar..156..570M, 2008ApJ...678..498N, 2008ApJ...686..580C}), and
Kozai cycles (with tidal friction) for planets in binary systems (e.g.,
\cite{2003ApJ...589..605W, 2005ApJ...627.1001T, 2007ApJ...669.1298F,
2007ApJ...670..820W}).  In particular, planet-planet scattering and the
Kozai effect may help to explain the eccentric orbits of close-in
planets, since Type II migration alone does not seem to produce large
orbital eccentricities.

Observations of transiting planets may lead to better understanding of
migration mechanisms through the Rossiter-McLaughlin effect (hereafter
the RM effect: \cite{1924ApJ....60...15R},
\cite{1924ApJ....60...22M}). The RM effect is an apparent
radial-velocity anomaly that is observed during planetary transits
(see \cite{2005ApJ...622.1118O, 2006ApJ...650..408G,
  2007ApJ...655..550G} for theoretical discussions). By measuring and
modeling this anomaly, one can learn the sky-projected angle between
the stellar spin axis and the planetary orbital axis, denoted by
$\lambda$. The angle between these two axes has been recognized as an
useful diagnostic to investigate planet migration histories. The basic
premises are that planets are formed with their orbital axes parallel
to the stellar rotation axis, that Type II migration maintains a small
spin-orbit alignment angle, while planet-planet scattering and the
Kozai effect may produce large spin-orbit misalignments (e.g.,
\cite{2008ApJ...678..498N, 2008ApJ...686..580C}), and that tidal
effects do not subsequently alter the spin-orbit angle
\citep{2009arXiv0902.4563B}.  To the extent that these premises
are valid, observations of the RM effect in transiting planets
allow us to test planet migration models. One may compare the
distribution of observed spin-orbit alignment angles with theoretical
predictions based on the scenarios of planet-planet scattering or
Kozai cycles with tidal friction (Fabrycky \& Winn 2009).

Measurement of the RM effect is especially interesting for transiting
planets in eccentric orbits (hereafter ``transiting eccentric
planets'' or TEPs), because the migration mechanisms that produce
eccentric orbits may also produce large spin-orbit misalignments.
The number of known TEPs is still small.  Observations of the RM effect
have been reported for a few cases, including HAT-P-2b
\citep{2007ApJ...665L.167W, 2008A&A...481..529L}, XO-3b
\citep{2008A&A...488..763H, 2009arXiv0902.3461W}, and the subject of
this paper, HD~17156b.

HD~17156b was discovered in a Doppler survey
(\cite{2007ApJ...669.1336F}; F07) and later found to transit its parent
star \citep{2007A&A...476L..13B}.  This planetary system stands out
among the transiting planets for its large orbital eccentricity
($e=0.68$) and relatively long orbital period (21 days).  The host star
is very bright ($V=8.2$), facilitating the detection of the RM effect
and many other follow-up observations.  For the HD~17156 planetary
system, previously Narita~et~al.~(2008; N08) reported a possible
spin-orbit misalignment $\lambda = 62^{\circ} \pm 25^{\circ}$, based on
radial-velocity data obtained at the Okayama Astrophysical Observatory
(OAO).  A more decisive conclusion, however, required further precise
observations with better statistics. Indeed Cochran~et~al.~(2008; C08)
and Barbieri~et~al.~(2008; B08) obtained new radial-velocity data (C08:
HET and HJST datasets, B08: TNG dataset), and claimed a close alignment,
$\lambda = 9.4^{\circ} \pm 9.3^{\circ}$ (C08) and $\lambda =
-4.8^{\circ} \pm 5.3^{\circ}$ (B08)\footnote{ Note that B08 used $\beta
(\equiv-\lambda)$ defined in \citet{2006ApJ...650..408G} for the
projected spin-orbit alignment angle.}.

We report here on our new observations with the Subaru 8.2m telescope,
covering a full transit of HD~17156b and providing the larger and more
precise radial-velocity dataset for the system around the transit phase.
Section~2 summarizes our Subaru spectroscopic observations and
radial-velocity dataset, and section~3 describes the analysis procedure
for the RM effect.  Section~4 presents and discusses the results for the
derived system parameters.  Section 5 summarizes the main findings of
the present paper.

\section{Subaru Spectroscopy and Radial Velocities}

We observed HD~17156 over 6.5 hr covering a full transit of HD~17156b
with the High Dispersion Spectrograph (HDS:
\cite{2002PASJ...54..855N}) on the Subaru 8.2m telescope on UT 2008
November 7.  We employed the standard I2a set-up of the HDS, covering
the wavelength range 4940~\AA\ $< \lambda <$ 6180~\AA\, and using the
Iodine gas absorption cell for precise radial velocity measurements.
The slit width of $0\farcs4$ yielded a spectral resolution of
$\sim$90000, and the seeing on that night was between $0\farcs4$ and
$0\farcs7$.  The exposure time was 3--5 minutes, yielding a typical
signal-to-noise ratio (SNR) of approximately 180 per pixel. We processed
the observed frames with standard IRAF\footnote{The Image Reduction
  and Analysis Facility (IRAF) is distributed by the U.S.\ National
  Optical Astronomy Observatories, which are operated by the
  Association of Universities for Research in Astronomy, Inc., under
  cooperative agreement with the National Science Foundation.}
procedures and extracted one-dimensional spectra.  We computed
relative radial velocities following the algorithm of
\citet{1996PASP..108..500B} and \citet{2002PASJ...54..873S}, as
described in \citet{2007PASJ...59..763N}.  We estimated the internal
error of each radial velocity as the scatter in the radial-velocity
solutions among the $\sim$4~\AA~segments of the spectrum.  The typical
internal error is 3~m~s$^{-1}$.  The radial velocities and their
internal errors are plotted in Figure~1 and summarized in Table~1.

\section{Analysis Procedure}

We empirically model the RM effect of HD17156 following the
procedure of Winn~et~al.~(2005). Specifically,
we simulate spectra that are affected by the RM effect, and then we
``measure'' the anomalous radial velocity of the simulated spectrum
using the same algorithm that we use on the actual data. We start with
the NSO solar spectrum \citep{1984sfat.book.....K} and deconvolve it
to remove the rotational broadening of the Sun. We neglect the
differential rotation of the Sun and adopt 1.85~km~s$^{-1}$ (the
rotational velocity at the equator) for the deconvolution kernel.  We
then apply a rotational broadening kernel with $V \sin I_s =
2.6$~km~s$^{-1}$ as appropriate to mimic the disk-integrated spectrum
of HD~17156 (F07).  In this step, for simplicity, we neglect
differential rotation, microturbulence and macroturbulence, convection
cells, and possible starspots and other active regions.  Finally, to
simulate the RM effect, we subtract a scaled and velocity-shifted copy
of the original unbroadened spectrum, representing the portion of the
stellar disk hidden by the planet.  We produce many simulated spectra
using different values of the scaling factor $f$ and the velocity
shift $v_p$, and compute the apparent radial velocity of each
spectrum.  We thereby determine an empirical relation between the size
of the planet and its position on the stellar disk, and the apparent
radial velocity of the star.  We fit a polynomial function to this
relation, and find
\begin{equation}
\Delta v = - f v_p \left[1.314 - 0.304
\left( \frac{v_p}{V \sin I_s} \right)^2 \right].
\end{equation}
As has been found previously (e.g., Winn et al.~2005), this formula
differs from the analytic formula presented by
\citet{2005ApJ...622.1118O}. The two formulas are nearly consistent,
however, if we rescale the value of $V \sin I_s$ in the analytic OTS
formula by a factor of 1.2.  In this sense the OTS formula may
overestimate the projected stellar rotational velocity by $\sim$20\%.
Further discussions for the empirical fitting formulae for the RM effect
are presented in section~4.3.

We analyze the RM effect of HD~17156b using our new Subaru dataset,
accompanied by published out-of-transit radial velocity datasets
presented in F07 (Subaru) and Winn~et~al.~(2009a; W09) (Keck).
(The Keck dataset was originally taken by F07, and W09 refined
the radial velocities and added new data.)
The out-of-transit data are needed to determine the
Keplerian orbital parameters of HD~17156b.  After performing this
analysis of the new transit data, we also repeated the analysis using
all of the published radial-velocity datasets around the transit phase
(N08, C08, and B08) for comparison.  In all of our analyses we
consider the error in each radial velocity data point to be the quadrature
sum of the internal error and 3~m~s$^{-1}$ (representing possible
systematic error such as ``stellar jitter'').

For radial velocity fitting, we fix stellar and planetary parameters
based on \citet{2008arXiv0812.0785B} as follows; the stellar mass $M_s
= 1.24$ [$M_{\odot}$], the stellar radius $R_s = 1.44$ [$R_{\odot}$],
the radius ratio $R_p/R_s = 0.0727$, and the semi-major axis $a =
0.1614$ [AU].  We also fix the quadratic limb-darkening parameters for
the spectroscopic band containing many Iodine absorption lines to $u_1
= 0.39$ and $u_2 = 0.37$ based on the tables of
\citet{2004A&A...428.1001C}. In addition, we adopt an updated transit
ephemeris based on simultaneous transit photometry as $T_c =
2454777.94761$ [HJD] and $P = 21.2165298$~days based on unpublished
data from the Transit Light Curve project (see e.g.\ Holman et
al.~2006, Winn et al.~2007). The adopted parameters are summarized in
Table~2.

Our model has 6 free parameters describing the HD~17156 system:
the radial velocity semiamplitude $K$, the eccentricity $e$,
the argument of periastron $\omega$,
the orbital inclination $i$,
the sky-projected stellar rotational velocity $V \sin I_s$,
and the sky-projected angle between the stellar spin axis and
the planetary orbital axis $\lambda$.
Our model also has one free parameter, a velocity offset,
for each independent radial velocity dataset
($v_1$: our Subaru dataset, $v_2$: Subaru in F07, $v_3$: Keck in W09,
$v_4$: HET in C08, $v_5$: HJST in C08, $v_6$: OAO in N08,
$v_7$: TNG in B08).

We calculate the $\chi^2$ statistic
\begin{eqnarray}
\chi^2 &=& \sum_i \left[ \frac{v_{i,{\rm obs}}-v_{i,{\rm calc}}}
{\sigma_{i}} \right]^2,
\end{eqnarray}
where
$v_{i, {\rm obs}}$ are the observed radial velocity data
and $v_{i, {\rm calc}}$ are the values calculated based on a Keplerian
orbit and on the RM calibration formula given above.
We determine optimal orbital parameters by minimizing the $\chi^2$
statistic using the AMOEBA algorithm \citep{1992nrca.book.....P},
and estimate confidence levels based on the criterion $\Delta \chi^2 = 1.0$
when a parameter is stepped away from its optimal value and the
other parameters are re-optimized.

\section{Results and Discussions}

\subsection{Results for the Key Parameter $\lambda$}

We first fit our Subaru dataset with the out-of-transit Subaru (F07)
and Keck (W09) datasets.
Figure~2 shows our Subaru dataset plotted with the best-fitting model curve.
We find $\lambda = 10.0^{\circ} \pm 5.1^{\circ}$,
indicating a fairly good alignment between the sky projections
of the stellar spin axis and the planetary orbital axis.
This value is similar to the result quoted by C08
($\lambda = 9.4^{\circ} \pm 9.3^{\circ}$), which was based on
F07, N08, and C08 datasets.

Next we conduct another analysis using all of the published datasets
(namely, with F07, N08, C08, B08, and W09 datasets).
Figure~3 plots all radial velocity data with the best-fitting model curve.
In this case we find $\lambda = 0.8^{\circ} \pm 4.3^{\circ}$,
and the results for the other parameters are in agreement
with the previous analysis.
The results of both analyses are summarized in Table~3.
We note that the central value of $\lambda$ for the combined
analysis is somewhat biased by the TNG dataset.
This is because the contribution to
$\chi^2$ from the TNG dataset is disproportionate ($\Delta\chi^2 =
55.73$ from 28 TNG radial velocities), whereas the ratio of
$\Delta\chi^2$ contribution and the number of data points is
approximately unity for all of the other datasets.  This suggests that
the errors in the TNG dataset may have been underestimated.  For this
reason the results of the joint analysis should be interpreted with
caution.

In addition, we conduct the same fitting procedure for the OAO dataset;
for the HET, HJST, and OAO datasets; and for the TNG dataset alone,
in order to make comparisons with previous studies
(N08, C08, and B08).
Note that we employ the same assumptions as described in section~3,
and we incorporate the out-of-transit velocities from
the F07/Subaru and W09/Keck datasets.
The results are $\lambda = 57^{\circ} \pm 23^{\circ}$
(as compared to $\lambda = 62^{\circ} \pm 25^{\circ}$ by N08)
for the OAO dataset;
$\lambda = 11.0^{\circ} \pm 11.8^{\circ}$
($\lambda = 9.4^{\circ} \pm 9.3^{\circ}$ by C08)
for HET, HJST, and OAO datasets; and
$\lambda = -33.7^{\circ} \pm 9.7^{\circ}$
($\lambda = -4.8^{\circ} \pm 5.3^{\circ}$ by B08)
for the TNG dataset.
Consequently, we confirm that the OAO dataset indeed implies
a large spin-orbit misalignment, but its poorer radial velocity accuracy
and time-resolution would have caused the fluke with lower
statistical significance.
In addition, although our analyses are well consistent with N08 and C08,
we are not able to reproduce the B08 result.
The systematic difference may be caused by the fact that B08
first fitted only the out-of-transit data and determined a radial velocity
offset (namely, the baseline of the Keplerian motion)
of the TNG dataset using fewer samples than we have here.
Such a treatment might be problematic since the RM amplitude
of HD~17156b is only $\sim15$~m~s$^{-1}$, and a difference
of offset velocity as much as several m~s$^{-1}$ would cause
a large systematic difference in $\lambda$.

\subsection{Assessments of Some Systematic Errors}

In the above analyses, we fixed some parameters
as summarized in Table~2, which were independently determined
by previous high accuracy photometric and spectroscopic studies.
In order to check the robustness of our results and to estimate
the level of systematic errors, we retry the fitting
for our Subaru dataset, the F07 Subaru dataset, and the
W09 Keck dataset with different adopted parameter values as follows;
$u_1 = 0.59$ (a greater limb-darkening case);
$u_1 = 0.19$ (a smaller limb-darkening case);
$a / R_s = 21.1$ (1$\sigma$ lower limit in W09); and
$a / R_s = 25.0$ (1$\sigma$ upper limit in W09).
We note that we find $R_p / R_s$ within its error reported in
B08 and W09 does not make a significant difference in results
(only $V \sin I_s$ is affected, and only by a few percent).

We first find that $K$, $e$, $\omega$, and offset velocities are
insensitive to the fixed parameters.
In addition, respective results for the spin-orbit alignment angle
$\lambda$ are
$10.8^{\circ} \pm 5.6^{\circ}$ ($u_1 = 0.59$);
$ 9.9^{\circ} \pm 5.0^{\circ}$ ($u_1 = 0.19$);
$ 5.8^{\circ} \pm 2.9^{\circ}$ ($a/R_s = 21.1$); and
$13.4^{\circ} \pm 7.0^{\circ}$ ($a/R_s = 25.0$).
Consequently, we confirm all results of $\lambda$ for the test cases
are well within our 1$\sigma$ uncertainty.

In contrast, we find appreciable systematic errors in the inclination
$i$ and $V \sin I_s$.
Results for the inclination $i$ are
$87.52^{\circ} \pm 0.36^{\circ}$ ($u_1 = 0.59$);
$86.99^{\circ} \pm 0.31^{\circ}$ ($u_1 = 0.19$);
$85.11^{\circ} \pm 0.21^{\circ}$ ($a/R_s = 21.1$); and
$87.95^{\circ} \pm 0.40^{\circ}$ ($a/R_s = 25.0$).
We thus conclude that the dependence of the inclination on the
limb-darkening parameters is small (within $\sim1\sigma$ error), but
the dependence on $a/R_s$ is as large as a few degrees (over 3$\sigma$).
When the covariance with the $a/R_s$ parameter is taken into account,
our constraint on $i$ based on the RM anomaly is approximately
$84.90^{\circ}$--$88.35^{\circ}$.
This is in good agreement with the constraint by
W09 ($85.4^{\circ}$--$88.3^{\circ}$).
On the other hand, we derive the projected velocity of
stellar rotation $V \sin I_s$ as,
$4.08 \pm 0.32 $~km~s$^{-1}$ ($u_1 = 0.59$);
$4.18 \pm 0.31 $~km~s$^{-1}$ ($u_1 = 0.19$);
$5.08 \pm 0.41 $~km~s$^{-1}$ ($a/R_s = 21.1$); and
$4.02 \pm 0.30 $~km~s$^{-1}$ ($a/R_s = 25.0$).
Based on these results we find that the result for $V \sin I_s$
is more robust than that for $i$, but the true error is
approximately 1~km~s$^{-1}$ (the span of the results just quoted).

We also consider possible
time-correlated errors in the radial velocity datasets, due to
possible instrumental or astrophysical effects. When present, this
so-called ``red noise'' complicates the accurate estimation of
parameter errors \citep{2006MNRAS.373..231P}.  To check for red noise
in the Subaru dataset, we calculate the standard deviation of the
residuals between the data and the best-fitting model, and also the
time-averaged (binned) residuals.  In the absence of time-correlated
noise, the standard deviation of $n$-point binned residuals
($\sigma_n$) would decrease as
\begin{equation}
\sigma_n = \sigma_1 \left(
\frac{1}{n} \frac{m}{m-1}
\right)^{1/2},
\end{equation}
where $m$ is the number of binned residuals.  We find the standard
deviation of residuals to be $\sigma_1 = 3.42$~m~s$^{-1}$, and if
there is no correlated noise then we expect the standard deviation of
5-point binned residuals to be $\sigma_{5,\rm{exp}} =
1.58$~m~s$^{-1}$.  The actual standard deviation of residuals with
5-point binning is $\sigma_{5,\rm{act}} = 1.69$~m~s$^{-1}$.  This is
only 7\% larger than the expected value; apparently the level of
time-correlated noise is small.

\subsection{Notes on the RM Formulae}

\citet{2005ApJ...631.1215W} pointed out a systematic difference in the
amplitude of the RM anomaly between the results of RM simulations and
the analytic OTS formula for the HD~209458 system.  A preliminary
analysis of Hirano~et~al. (in~prep.) indicates the systematic
difference between the OTS formula and the RM simulations is due to the
fact that the former computes the spectral shift via intensity-weighted
average while the latter via the cross-correlaiton. Specifically, they
showed that the amplitude of the systematics depends on the spin rotation
velocity of the host star.  Following their procedure, we also
confirmed the similar systematics for the HD~17156 system using
the Subaru analysis routine (section~3). In addition to that,
we show in this subsection that the input value of the spin rotation
velocity, more specifically $V \sin I_s$, in the empirical RM formula
does not bias the estimate of the other parameters.

When we fit our Subaru dataset using the OTS formula, we find $V \sin
I_s = 5.03$~km~s$^{-1}$, which is $\sim$20\% higher than the result using
the simulation-based formula.  The results for all the other parameters
are essentially unchanged.  Thus the choice of the RM calibration
formula corresponds to a $\sim$20\% change in the inferred value of $V
\sin I_s$.  We also expect that the formula of
\citet{2006ApJ...650..408G} would give similar results to the OTS
formula, as both formulas were calculated under the same assumption:
that the anomalous radial velocity is equal to the shift in the
intensity-weighted mean wavelength of the absorption lines.

In the previous studies, the $V \sin I_s$ value assumed for the
simulation was generally in agreement with the resultant fitted
value. However, we obtained $V \sin I_s = 4.18 \pm 0.31$~km~s$^{-1}$
from the RM analyses even though we assumed $V \sin I_s =
2.6$~km~s$^{-1}$ in the simulations.  Ideally the fitted value of $V
\sin I_s$ should be consistent with the assumed one.  In order to check
the dependence of assumed $V \sin I_s$ value on fitting results, we
reran the simulations assuming $V \sin I_s = 4.2$~km~s$^{-1}$ for the
convolution kernel instead of the original choice of 2.6~km~s$^{-1}$.
The revised empirical formula is
\begin{equation}
\Delta v = - f v_p \left[1.367 - 0.505
\left( \frac{v_p}{V \sin I_s} \right)^2 \right].
\end{equation}
With this revised formula, we repeated the analysis using the Subaru
dataset along with the F07 and W09 datasets.  We find $V \sin I_s = 4.22
\pm 0.31$~km~s$^{-1}$ and the values of the other parameters are
essentially unchanged. The results are now self-consistent and
suggest that the choice of an empirical formula does not make a
significant difference. Thus we conclude that the determination of
$\lambda$ of the current system is fairly insensitive to the choice
of the RM formula.

\section{Summary}

We observed a full transit of HD~17156b with the Subaru 8.2m telescope,
and measured radial velocities with the highest time-resolution and
accuracy that have yet been presented for transit observations of this
system.  We analyzed the RM effect and found that the sky-projected
spin-orbit alignment angle of the HD~17156 system is $\lambda =
10.0^{\circ} \pm 5.1^{\circ}$. This result supersedes the previous claim
of $\lambda = 62^{\circ} \pm 25^{\circ}$ by N08, which was due to lower
precision and poor statistics of the previous datasets. Thus we conclude
that the HD~17156 is not strongly misaligned, and as such it joins the
majority of other systems for which the RM effect has been detected
(\cite{2009arXiv0902.0737F} and references therein).  If the projected
spin-orbit angle is representative of the true spin-orbit angle, then
the orbital tilt of HD~17156b is not so different from the orbital tilt
of the Jupiter relative to the equatorial plane of the Sun.

The other TEPs for which RM observations have been reported include
HAT-P-2b and XO-3b.  Interestingly the outcomes of the measurements of
the two systems were quite different; HAT-P-2b is also well-aligned
\citep{2007ApJ...665L.167W, 2008A&A...481..529L} while XO-3b has a
significant spin-orbit misalignment \citep{2008A&A...488..763H,
  2009arXiv0902.3461W}.  The existence of at least some
highly-inclined planets would support the notion that some TEPs have
migrated through planet-planet scattering or the Kozai effect and
tidal friction.  However, with just three samples one cannot draw
robust conclusions (see e.g.\ \cite{2009arXiv0902.0737F}).  Further
observations of the RM effect for TEPs are desired. Recently the
highly eccentric, long period planet HD~80606b has been added to the
list of TEPs.  Observations of the RM effect for that system were
presented by \citet{2009arXiv0902.4457M}, but because of their limited time
coverage a robust determination of $\lambda$ is not yet possible.  It
will be very interesting to see the rest of the RM anomaly filled in
with future observations.

We acknowledge invaluable support of our observations by Akito Tajitsu,
a support scientist for the Subaru HDS.  This paper is based on data
collected at Subaru Telescope, which is operated by the National
Astronomical Observatory of Japan.  The data analysis was in part
carried out on common use data analysis computer system at the Astronomy
Data Center, ADC, of the National Astronomical Observatory of Japan.
N.N. is supported by a Japan Society for Promotion of Science (JSPS)
Fellowship for Research (PD: 20-8141).
This work was partly supported by World Premier
International Research Center Initiative (WPI Initiative), MEXT, Japan.
We appreciate a careful reading and quick comments by the referee,
William Cochran.
Finally, we wish to acknowledge the very significant cultural role
and reverence that the summit of Mauna Kea has always had within
the indigenous Hawaiian community.

\begin{figure}[pthb]
 \begin{center}
  \FigureFile(130mm,130mm){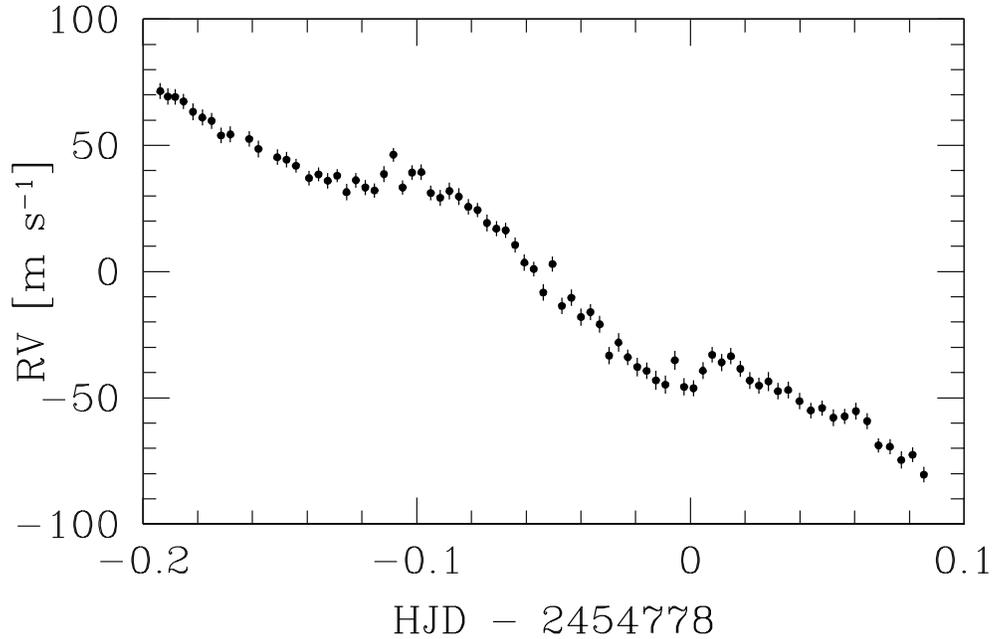}
 \end{center}
  \caption{Radial velocities taken with the Subaru HDS.
  The values and errors are presented in table~1.
  \label{ourdata}}
\end{figure}

\begin{figure}[pthb]
 \begin{center}
  \FigureFile(130mm,130mm){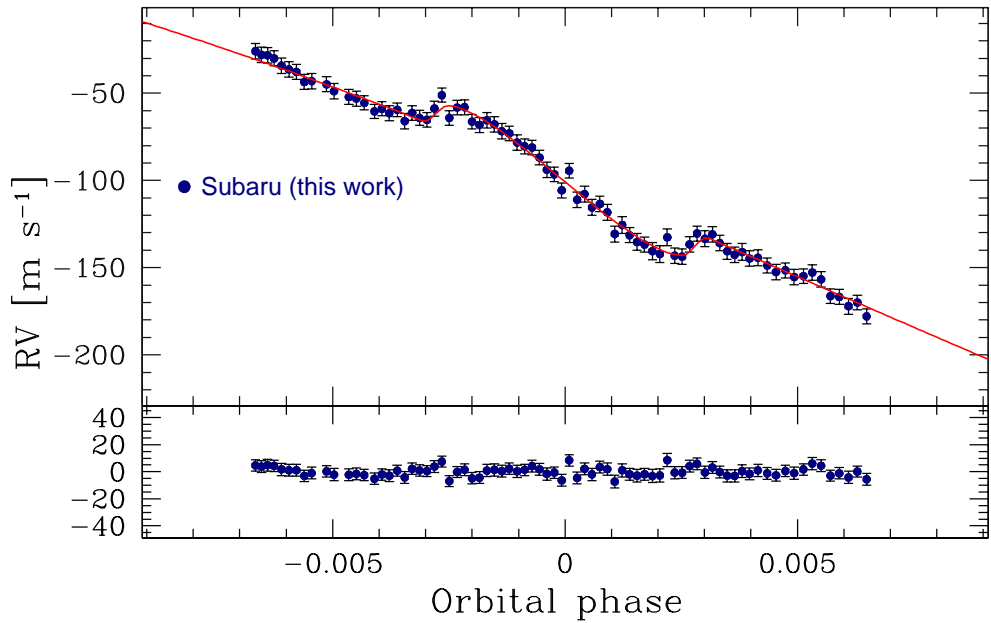}
 \end{center}
  \caption{
  The upper panel:
  Radial velocities and the best-fit curve of HD~17156 as a function of
  orbital phase. Only Subaru and Keck datasets are used.
  The lower panel: Residuals from the best-fit curve.
  \label{bestfitsubaru}}
\end{figure}

\begin{figure*}[pthb]
 \begin{center}
  \FigureFile(160mm,160mm){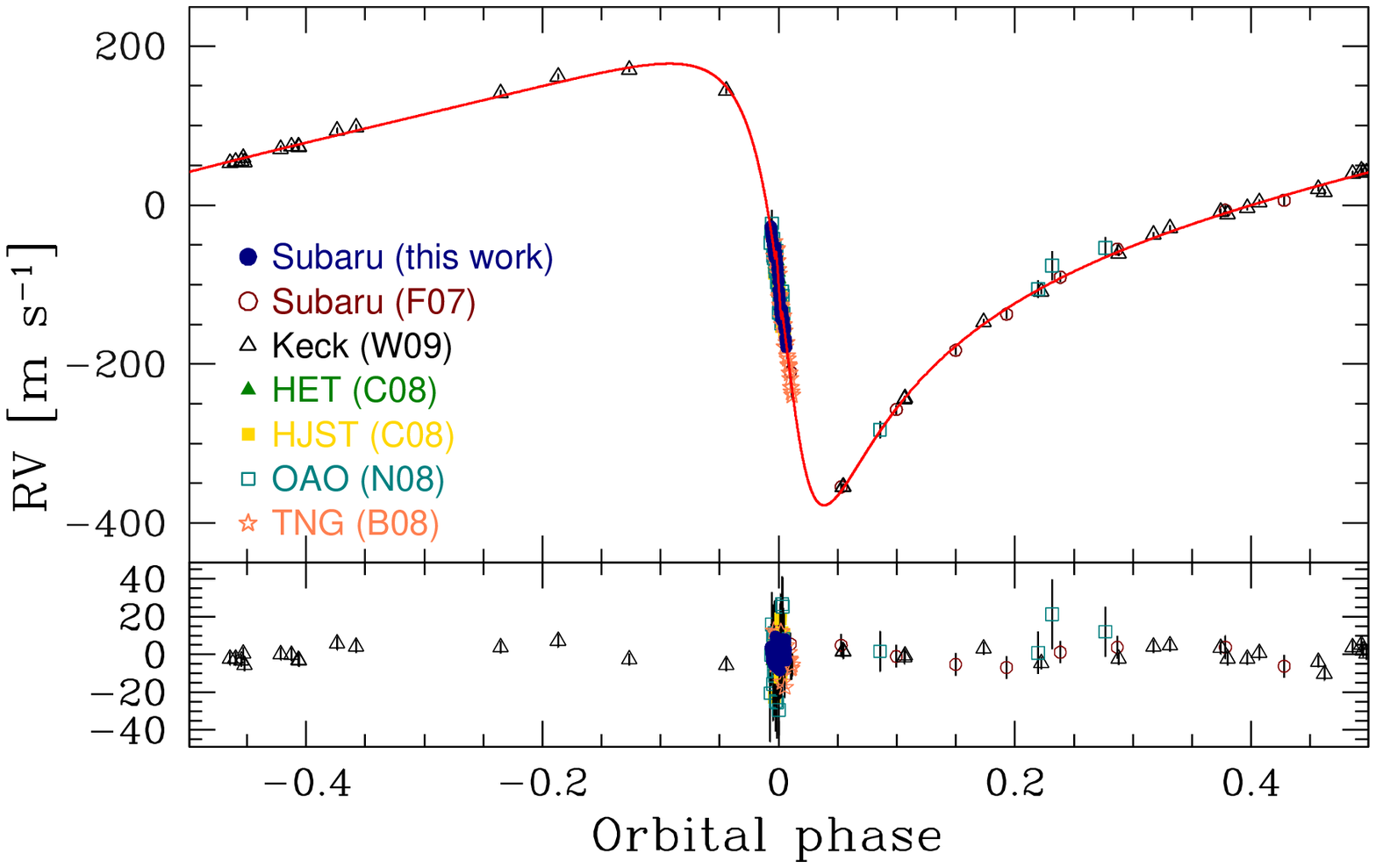}\hspace{5mm}
  \FigureFile(160mm,160mm){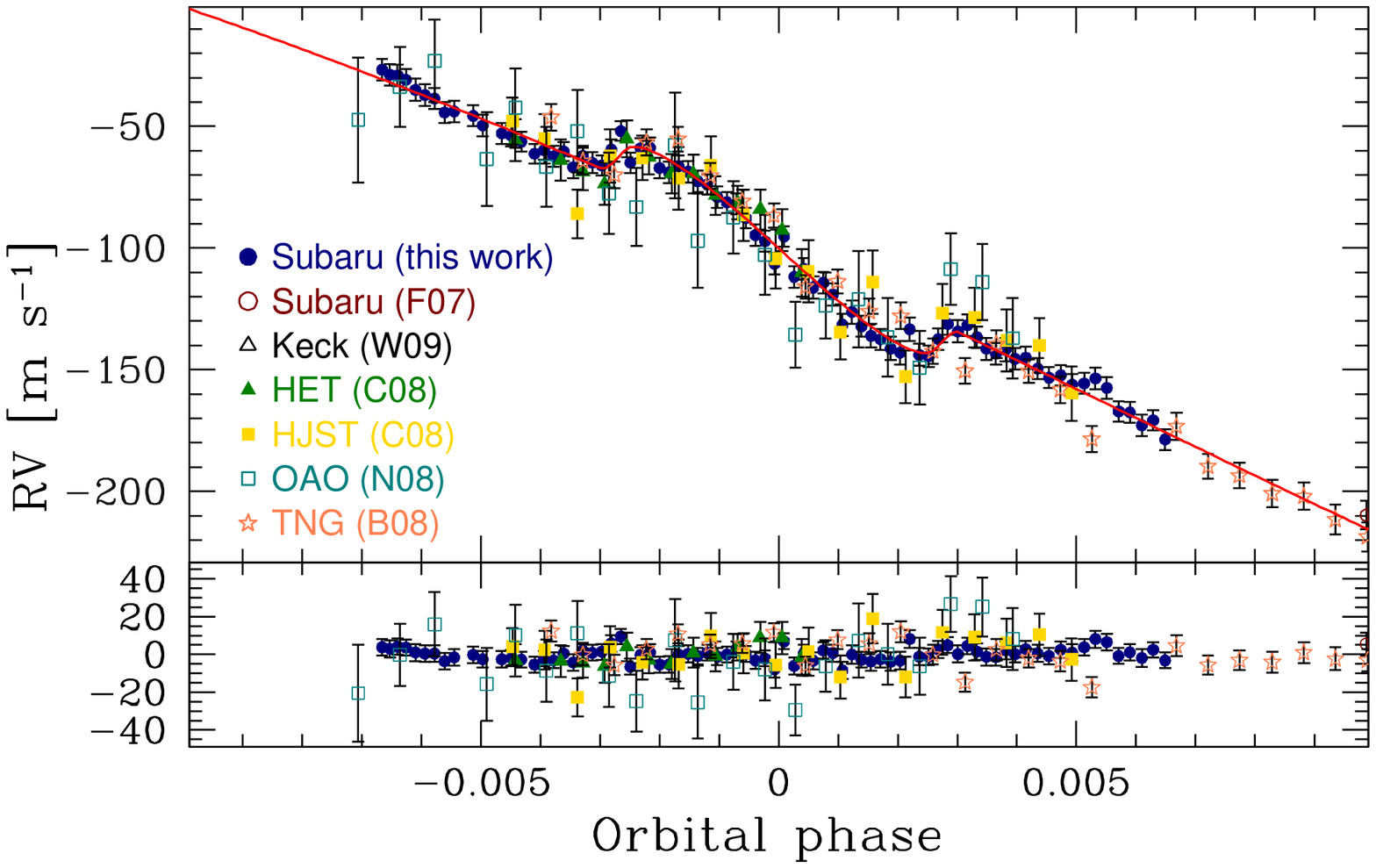}
 \end{center}
  \caption{
  The upper panels:
  All published radial velocities and the best-fit curve of HD~17156
  as a function of orbital phase. The upper figure shows the entire orbit
  and the lower figure does the zoom of transit phase.
  The lower panels: Residuals from the best-fit curve.
  \label{bestfitall}}
\end{figure*}

\begin{longtable}[c]{lcc}
\caption{Radial velocities obtained with the Subaru/HDS.}
\hline
\endhead
\hline
\endfoot
\hline
Time [HJD]  & Value [m~s$^{-1}$] & Error [m~s$^{-1}$]\\
\hline
2454777.80631 &	71.48 &	3.15 \\
2454777.80904 &	69.36 &	3.26 \\
2454777.81176 &	69.19 &	3.15 \\
2454777.81483 &	67.40 &	3.06 \\
2454777.81825 &	63.25 &	3.34 \\
2454777.82166 &	61.08 &	3.15 \\
2454777.82508 &	59.66 &	3.13 \\
2454777.82850 &	53.92 &	3.10 \\
2454777.83193 &	54.38 &	3.16 \\
2454777.83877 &	52.56 &	3.12 \\
2454777.84219 &	48.59 &	3.33 \\
2454777.84903 &	45.32 &	3.07 \\
2454777.85245 &	44.35 &	3.13 \\
2454777.85587 &	41.94 &	2.81 \\
2454777.86066 &	37.00 &	2.87 \\
2454777.86409 &	38.47 &	2.73 \\
2454777.86751 &	35.98 &	3.04 \\
2454777.87093 &	37.97 &	2.59 \\
2454777.87435 &	31.49 &	3.26 \\
2454777.87778 &	36.16 &	2.88 \\
2454777.88120 &	33.32 &	3.05 \\
2454777.88461 &	32.15 &	2.82 \\
2454777.88803 &	38.62 &	3.14 \\
2454777.89145 &	46.30 &	2.74 \\
2454777.89487 &	33.32 &	2.88 \\
2454777.89829 &	39.22 &	2.85 \\
2454777.90171 &	39.40 &	3.10 \\
2454777.90514 &	31.14 &	2.88 \\
2454777.90856 &	29.22 &	3.20 \\
2454777.91198 &	31.93 &	3.30 \\
2454777.91540 &	29.64 &	3.33 \\
2454777.91882 &	25.67 &	3.11 \\
2454777.92224 &	24.40 &	2.81 \\
2454777.92566 &	19.20 &	3.32 \\
2454777.92909 &	17.00 &	3.01 \\
2454777.93251 &	16.33 &	3.02 \\
2454777.93594 &	10.53 &	3.02 \\
2454777.93936 &	3.58 &	3.31 \\
2454777.94278 &	1.02 &	2.93 \\
2454777.94621 &	-8.27 &	3.19 \\
2454777.94962 &	3.00 &	2.98 \\
2454777.95304 &	-13.59 &	3.21 \\
2454777.95647 &	-10.38 &	3.28 \\
2454777.96002 &	-18.02 &	3.43 \\
2454777.96345 &	-16.04 &	3.15 \\
2454777.96687 &	-20.84 &	3.32 \\
2454777.97029 &	-33.28 &	3.48 \\
2454777.97371 &	-28.10 &	3.72 \\
2454777.97714 &	-33.93 &	3.10 \\
2454777.98057 &	-37.80 &	3.73 \\
2454777.98399 &	-39.30 &	3.19 \\
2454777.98743 &	-43.08 &	3.79 \\
2454777.99085 &	-44.78 &	3.67 \\
2454777.99428 &	-35.16 &	3.71 \\
2454777.99770 &	-45.68 &	3.41 \\
2454778.00112 &	-46.25 &	3.19 \\
2454778.00454 &	-39.24 &	3.35 \\
2454778.00796 &	-32.99 &	3.06 \\
2454778.01139 &	-36.00 &	3.40 \\
2454778.01481 &	-33.47 &	3.20 \\
2454778.01823 &	-38.50 &	3.23 \\
2454778.02165 &	-43.15 &	3.41 \\
2454778.02508 &	-45.17 &	3.21 \\
2454778.02851 &	-43.56 &	3.72 \\
2454778.03194 &	-47.37 &	3.24 \\
2454778.03570 &	-46.86 &	3.35 \\
2454778.03983 &	-51.32 &	3.22 \\
2454778.04396 &	-55.03 &	3.15 \\
2454778.04809 &	-54.08 &	3.04 \\
2454778.05222 &	-57.89 &	3.31 \\
2454778.05635 &	-57.39 &	3.01 \\
2454778.06048 &	-55.33 &	3.39 \\
2454778.06461 &	-59.25 &	3.23 \\
2454778.06874 &	-68.85 &	2.81 \\
2454778.07288 &	-69.28 &	3.00 \\
2454778.07701 &	-74.63 &	3.42 \\
2454778.08114 &	-72.57 &	2.90 \\
2454778.08527 &	-80.46 &	3.12 \\
\hline
\label{rvsummary}
\end{longtable}

\begin{table}[t]
\caption{Adopted stellar and planetary parameters.}
\begin{center}
\begin{tabular}{l|cc}
\hline
Parameter & Value & Source \\
\hline
$M_s$ [$M_{\odot}$] 
& $1.24$ & B08 \\
$R_s$ [$R_{\odot}$]
& $1.44$ & B08 \\
$R_p/R_s$ 
& $0.0727$ & B08 \\
$a$ [AU]
& $0.1614$ & B08 \\
$u_1$
& $0.39$ & \citet{2004A&A...428.1001C} \\
$u_2$
& $0.37$ & \citet{2004A&A...428.1001C} \\
$T_c$ [HJD]
& $2454777.94761$ & Winn et al. in prep. \\
$P$ [days]
& $21.2165298$ & Winn et al. in prep. \\
\hline
\multicolumn{3}{l}{\hbox to 0pt{\parbox{80mm}{\footnotesize
}\hss}}
\end{tabular}
\label{adoptedpar}
\end{center}
\end{table}

\begin{table}[t]
\caption{Fitted values and uncertainties of the free parameters.}
\begin{center}
\begin{tabular}{l|cc|cc}
\hline
 & \multicolumn{2}{c|}{Subaru and Keck}
 & \multicolumn{2}{c}{all datasets} \\
Parameter & Value & Uncertainty & Value & Uncertainty \\
\hline
$K$ [m s$^{-1}$] 
& 275.99  & $\pm 1.27$
& 277.77  & $\pm 1.19$\\
$e$ 
& 0.6801  & $\pm 0.0019$
& 0.6835  & $\pm 0.0017$\\
$\omega$ [$^{\circ}$]
& 121.62  & $\pm 0.42$
& 121.73  & $\pm 0.41$\\
$i$ [$^{\circ}$]\,\,$^{\rm{a}}$
& 87.21  & $\pm 0.31$
& 86.95  & $\pm 0.27$\\
$V \sin I_s$ [km s$^{-1}$]\,\,$^{\rm{a}}$
& 4.18  & $\pm 0.31$
& 4.07  & $\pm 0.28$\\
$\lambda$ [$^{\circ}$]
& 10.0  & $\pm 5.1$
&  0.8  & $\pm 4.3$\\
$v_1$ [m s$^{-1}$] 
& 97.46  & $\pm 1.57$
& 98.23  & $\pm 1.51$\\
$v_2$ [m s$^{-1}$] 
& 92.89  & $\pm 2.10$
& 93.12  & $\pm 2.09$\\
$v_3$ [m s$^{-1}$] 
& -0.11  & $\pm 0.64$
& -0.20  & $\pm 0.64$\\
$v_4$ [m s$^{-1}$] 
&   &   
& 74.21  & $\pm 2.71$\\
$v_5$ [m s$^{-1}$] 
&   &   
& -7850.55  & $\pm 3.01$\\
$v_6$ [m s$^{-1}$] 
&   &   
& 142.34 & $\pm 3.29$\\
$v_7$ [m s$^{-1}$] 
&   &   
& 146.68 & $\pm 1.89$\\
\hline
\multicolumn{5}{l}{\hbox to 0pt{\parbox{100mm}{\footnotesize
$v_1 \sim v_7$ are offset velocites for respective datasets
(see Section~3).\\
\footnotemark[a]: Systematic errors are not included in the uncertainties
(see Section~4.2).\\
}\hss}}
\end{tabular}
\label{resultpar}
\end{center}
\end{table}

\clearpage



\begin{thebibliography}{}
\expandafter\ifx\csname natexlab\endcsname\relax\def\natexlab#1{#1}\fi

\bibitem[{{Barbieri} {et~al.}(2007){Barbieri}, {Alonso}, {Laughlin},
  {Almenara}, {Bissinger}, {Davies}, {Gasparri}, {Guido}, {Lopresti},
  {Manzini}, \& {Sostero}}]{2007A&A...476L..13B}
{Barbieri}, M., et al. 2007, \aap, 476, L13

\bibitem[{{Barbieri} {et~al.}(2008){Barbieri}, {Alonso}, {Desidera},
  {Sozzetti}, {Martinez Fiorenzano}, {Almenara}, {Cecconi}, {Claudi},
  {Charbonneau}, {Endl}, {Granata}, {Gratton}, {Laughlin}, \&
  {Loeillet}}]{2008arXiv0812.0785B}
{Barbieri}, M., et al. 2008, ArXiv e-prints (B08)

\bibitem[{{Barker} \& {Ogilvie}(2009)}]{2009arXiv0902.4563B}
{Barker}, A.~J. \& {Ogilvie}, G.~I. 2009, ArXiv e-prints

\bibitem[{{Butler} {et~al.}(1996){Butler}, {Marcy}, {Williams}, {McCarthy},
  {Dosanjh}, \& {Vogt}}]{1996PASP..108..500B}
{Butler}, R.~P., {Marcy}, G.~W., {Williams}, E., {McCarthy}, C., {Dosanjh}, P.,
  \& {Vogt}, S.~S. 1996, \pasp, 108, 500

\bibitem[{{Chatterjee} {et~al.}(2008){Chatterjee}, {Ford}, {Matsumura}, \&
  {Rasio}}]{2008ApJ...686..580C}
{Chatterjee}, S., {Ford}, E.~B., {Matsumura}, S., \& {Rasio}, F.~A. 2008, \apj,
  686, 580

\bibitem[{{Claret}(2004)}]{2004A&A...428.1001C}
{Claret}, A. 2004, \aap, 428, 1001

\bibitem[{{Cochran} {et~al.}(2008){Cochran}, {Redfield}, {Endl}, \&
  {Cochran}}]{2008ApJ...683L..59C}
{Cochran}, W.~D., {Redfield}, S., {Endl}, M., \& {Cochran}, A.~L. 2008, \apjl,
  683, L59 (C08)

\bibitem[{{Fabrycky} \& {Tremaine}(2007)}]{2007ApJ...669.1298F}
{Fabrycky}, D.~C. \& {Tremaine}, S. 2007, \apj, 669, 1298

\bibitem[{{Fabrycky} \& {Winn}(2009)}]{2009arXiv0902.0737F}
{Fabrycky}, D.~C. \& {Winn}, J.~N. 2009, ArXiv e-prints

\bibitem[{{Fischer} {et~al.}(2007){Fischer}, {Vogt}, {Marcy}, {Butler}, {Sato},
  {Henry}, {Robinson}, {Laughlin}, {Ida}, {Toyota}, {Omiya}, {Driscoll},
  {Takeda}, {Wright}, \& {Johnson}}]{2007ApJ...669.1336F}
{Fischer}, D.~A., et al. 2007, \apj, 669, 1336 (F07)

\bibitem[{{Gaudi} \& {Winn}(2007)}]{2007ApJ...655..550G}
{Gaudi}, B.~S. \& {Winn}, J.~N. 2007, \apj, 655, 550

\bibitem[{{Gim{\'e}nez}(2006)}]{2006ApJ...650..408G}
{Gim{\'e}nez}, A. 2006, \apj, 650, 408

\bibitem[{{H{\'e}brard} {et~al.}(2008){H{\'e}brard}, {Bouchy}, {Pont},
  {Loeillet}, {Rabus}, {Bonfils}, {Moutou}, {Boisse}, {Delfosse}, {Desort},
  {Eggenberger}, {Ehrenreich}, {Forveille}, {Lagrange}, {Lovis}, {Mayor},
  {Pepe}, {Perrier}, {Queloz}, {Santos}, {S{\'e}gransan}, {Udry}, \&
  {Vidal-Madjar}}]{2008A&A...488..763H}
{H{\'e}brard}, G., et al. 2008, \aap,
  488, 763

\bibitem[{{Holman} {et~al.}(2006){Holman}, {Winn}, {Latham}, {O'Donovan},
  {Charbonneau}, {Bakos}, {Esquerdo}, {Hergenrother}, {Everett}, \&
  {P{\'a}l}}]{2006ApJ...652.1715H}
{Holman}, M.~J., et al. 2006, \apj, 652, 1715

\bibitem[{{Ida} \& {Lin}(2004)}]{2004ApJ...616..567I}
{Ida}, S. \& {Lin}, D.~N.~C. 2004, \apj, 616, 567

\bibitem[{{Kurucz} {et~al.}(1984){Kurucz}, {Furenlid}, {Brault}, \&
  {Testerman}}]{1984sfat.book.....K}
{Kurucz}, R.~L., {Furenlid}, I., {Brault}, J., \& {Testerman}, L. 1984,
  in {Solar flux atlas from 296 to 1300 nm}, ed. R.~L. {Kurucz}, I.~{Furenlid},
  J.~{Brault}, \& L.~{Testerman}

\bibitem[{{Lin} {et~al.}(1996){Lin}, {Bodenheimer}, \&
  {Richardson}}]{1996Natur.380..606L}
{Lin}, D.~N.~C., {Bodenheimer}, P., \& {Richardson}, D.~C. 1996, \nat, 380, 606

\bibitem[{{Lin} \& {Papaloizou}(1985)}]{1985prpl.conf..981L}
{Lin}, D.~N.~C. \& {Papaloizou}, J. 1985, in Protostars and Planets II, ed.
  D.~C. {Black} \& M.~S. {Matthews}, 981--1072

\bibitem[{{Loeillet} {et~al.}(2008){Loeillet}, {Shporer}, {Bouchy}, {Pont},
  {Mazeh}, {Beuzit}, {Boisse}, {Bonfils}, {da Silva}, {Delfosse}, {Desort},
  {Ecuvillon}, {Forveille}, {Galland}, {Gallenne}, {H{\'e}brard}, {Lagrange},
  {Lovis}, {Mayor}, {Moutou}, {Pepe}, {Perrier}, {Queloz}, {S{\'e}gransan},
  {Sivan}, {Santos}, {Tsodikovich}, {Udry}, \&
  {Vidal-Madjar}}]{2008A&A...481..529L}
{Loeillet}, B., et al. 2008,
  \aap, 481, 529

\bibitem[{{Marzari} \& {Weidenschilling}(2002)}]{2002Icar..156..570M}
{Marzari}, F. \& {Weidenschilling}, S.~J. 2002, Icarus, 156, 570

\bibitem[{{McLaughlin}(1924)}]{1924ApJ....60...22M}
{McLaughlin}, D.~B. 1924, \apj, 60, 22

\bibitem[{{Moutou} {et~al.}(2009){Moutou}, {Hebrard}, {Bouchy}, {Eggenberger},
  {Boisse}, {Bonfils}, {Gravallon}, {Ehrenreich}, {Forveille}, {Delfosse},
  {Desort}, {Lagrange}, {Lovis}, {Mayor}, {Pepe}, {Perrier}, {Pont}, {Queloz},
  {Santos}, {Segransan}, {Udry}, \& {Vidal-Madjar}}]{2009arXiv0902.4457M}
{Moutou}, C., et al. 2009, ArXiv e-prints

\bibitem[{{Nagasawa} {et~al.}(2008){Nagasawa}, {Ida}, \&
  {Bessho}}]{2008ApJ...678..498N}
{Nagasawa}, M., {Ida}, S., \& {Bessho}, T. 2008, \apj, 678, 498

\bibitem[{{Narita} {et~al.}(2007){Narita}, {Enya}, {Sato}, {Ohta}, {Winn},
  {Suto}, {Taruya}, {Turner}, {Aoki}, {Tamura}, {Yamada}, \&
  {Yoshii}}]{2007PASJ...59..763N}
{Narita}, N., et al. 2007, \pasj, 59, 763

\bibitem[{{Narita} {et~al.}(2008){Narita}, {Sato}, {Ohshima}, \&
  {Winn}}]{2008PASJ...60L...1N}
{Narita}, N., {Sato}, B., {Ohshima}, O., \& {Winn}, J.~N. 2008, \pasj, 60,
L1+ (N08)

\bibitem[{{Noguchi} {et~al.}(2002){Noguchi}, {Aoki}, {Kawanomoto}, {Ando},
  {Honda}, {Izumiura}, {Kambe}, {Okita}, {Sadakane}, {Sato}, {Tajitsu},
  {Takada-Hidai}, {Tanaka}, {Watanabe}, \& {Yoshida}}]{2002PASJ...54..855N}
{Noguchi}, K., et al. 2002, \pasj, 54, 855

\bibitem[{{Ohta} {et~al.}(2005){Ohta}, {Taruya}, \&
  {Suto}}]{2005ApJ...622.1118O}
{Ohta}, Y., {Taruya}, A., \& {Suto}, Y. 2005, \apj, 622, 1118

\bibitem[{{Pont} {et~al.}(2006){Pont}, {Zucker}, \&
  {Queloz}}]{2006MNRAS.373..231P}
{Pont}, F., {Zucker}, S., \& {Queloz}, D. 2006, \mnras, 373, 231

\bibitem[{{Press} {et~al.}(1992){Press}, {Teukolsky}, {Vetterling}, \&
  {Flannery}}]{1992nrca.book.....P}
{Press}, W.~H., {Teukolsky}, S.~A., {Vetterling}, W.~T., \& {Flannery}, B.~P.
  1992, {Numerical recipes in C. The art of scientific computing} (Cambridge:
  University Press, |c1992, 2nd ed.)

\bibitem[{{Rasio} \& {Ford}(1996)}]{1996Sci...274..954R}
{Rasio}, F.~A. \& {Ford}, E.~B. 1996, Science, 274, 954

\bibitem[{{Rossiter}(1924)}]{1924ApJ....60...15R}
{Rossiter}, R.~A. 1924, \apj, 60, 15

\bibitem[{{Sato} {et~al.}(2002){Sato}, {Kambe}, {Takeda}, {Izumiura}, \&
  {Ando}}]{2002PASJ...54..873S}
{Sato}, B., {Kambe}, E., {Takeda}, Y., {Izumiura}, H., \& {Ando}, H. 2002,
  \pasj, 54, 873

\bibitem[{{Takeda} \& {Rasio}(2005)}]{2005ApJ...627.1001T}
{Takeda}, G. \& {Rasio}, F.~A. 2005, \apj, 627, 1001

\bibitem[{{Winn} {et~al.}(2005){Winn}, {Noyes}, {Holman}, {Charbonneau},
  {Ohta}, {Taruya}, {Suto}, {Narita}, {Turner}, {Johnson}, {Marcy}, {Butler},
  \& {Vogt}}]{2005ApJ...631.1215W}
{Winn}, J.~N., et al. 2005, \apj, 631, 1215

\bibitem[{{Winn} {et~al.}(2007{\natexlab{a}}){Winn}, {Holman}, \&
  {Fuentes}}]{2007AJ....133...11W}
{Winn}, J.~N., {Holman}, M.~J., \& {Fuentes}, C.~I. 2007{\natexlab{a}}, \aj,
  133, 11

\bibitem[{{Winn} {et~al.}(2007{\natexlab{b}}){Winn}, {Johnson}, {Peek}, {Marcy}, {Bakos},
  {Enya}, {Narita}, {Suto}, {Turner}, \& {Vogt}}]{2007ApJ...665L.167W}
{Winn}, J.~N., et al. 2007{\natexlab{b}}, \apjl, 665, L167

\bibitem[{{Winn} {et~al.}(2009{\natexlab{a}}){Winn}, {Holman}, {Henry},
  {Torres}, {Fischer}, {Johnson}, {Marcy}, {Shporer}, \&
  {Mazeh}}]{2009ApJ...693..794W}
{Winn}, J.~N., et al. 2009{\natexlab{a}}, \apj, 693, 794 (W09)

\bibitem[{{Winn} {et~al.}(2009{\natexlab{b}}){Winn}, {Johnson}, {Fabrycky},
  {Howard}, {Marcy}, {Narita}, {Crossfield}, {Suto}, {Turner}, {Esquerdo}, \&
  {Holman}}]{2009arXiv0902.3461W}
{Winn}, J.~N., et al. 2009{\natexlab{b}}, ArXiv e-prints

\bibitem[{{Wu} \& {Murray}(2003)}]{2003ApJ...589..605W}
{Wu}, Y. \& {Murray}, N.~W. 2003, \apj, 589, 605

\bibitem[{{Wu} {et~al.}(2007){Wu}, {Murray}, \&
  {Ramsahai}}]{2007ApJ...670..820W}
{Wu}, Y., {Murray}, N.~W., \& {Ramsahai}, J.~M. 2007, \apj, 670, 820

\end{thebibliography}
\end{document}